# Instructional Level Parallelism

Taposh Dutta Roy, *Student Member, IEEE*

*Abstract*—**This paper is a review of the developments in Instruction level parallelism. It takes into account all the changes made in speeding up the execution. The various drawbacks and dependencies due to pipelining are discussed and various solutions to overcome them are also incorporated. It goes ahead in the last section to explain where is the new research leading us.**

## I. INTRODUCTION

IN order to improve the performance of the machine, parallelism or pipelining is used. The basic idea in using pipelining is to make use of multiple instructions in one clock cycle. This is possible only if there are no dependencies between the two instructions. In a nutshell, we can say that Instructional Level Parallelism is an issue that deals with exploiting parallelism at instructional level. It basically deals with the scheduling the instructions such that there are no hazards.

It is important to note that when a programmer writes a program, he assumes serial execution of the instruction. The role of the ILP processor is to take a serial program (i.e. a specification of operations to execute)remove much of the non-essential sequential execution in this specification so that the program can be processed in parallel, and turn the program into a high performance version. The above transformation must occur transparent to the programmer. An ILP processor that executes the serial program in parallel is called a Super-Scalar processor. The basic idea of this paper is to discuss the instructional level parallelism - concept, hazards, limitations and future trends.

## II. ILP AND ITS RELEVANCE TO COMPUTER ARCHITECTURE

Since the beginning of computing history, computer architects have been craving for speed i.e., improving the number of tasks to be implemented in a unit time. Instructional level Parallelism was obviously the most important break through in the technology. It paved the path for dealing with multiple instructions in one clock cycle. However there were many problems in putting to practice this simple idea. These problems were basically due to the dependencies between the instructions and the need of sequence.

This idea, first implemented in CDC 6600 computer, has greatly advanced and lot of ground has been captured in this respect. Instructional level parallelism is still an area of exploration and more work is expected in this field. In this paper, I plan to explain the concepts, the various drawbacks and strategies to solve them and future development.

## III. HISTORY OF INSTRUCTIONAL LEVEL PARALLELISM

### A. Initial Stage

The concept of *instructional level parallelism* was first developed in 1940s. The first idea of parallelism or horizontal micro-code came into print in Turning's 1946 design of the Pilot ACE and was carefully described by Wilkes [1951]. A major statement that is of historical importance is "*In some cases it may be possible for two or more micro-operations to take place at the same time*"[Wilkes and Stringer 1953]. During 1960s transistorized computers were prevalent, but with the advent of CDC 6600, a Control Data Corporation machine, basic instructions came into use. This paved the path for a more complex machine and at the right time IBM launched its 360/91, which also introduced instructional level parallelism. This machine based partly on IBM's instruction-level parallelism experiment offered less parallelism than the CDC 6600, but was far more ambitious than the CDC 6600 in its attempt to rearrange the instruction stream to keep the functional units busy. In the late 1970s the beginning of a new style of ILP, called *very long instruction word* (VLIW), came into picture. It could be seen as if the VLIW architectures were the natural growth towards parallel architecture. The VLIW architecture was prevalent in the 1980s and few computer startups started developing their computers with various levels of parallelism.

### B. Development Stage

In the 1990s, with the development in silicon technology, RISC computers were developed with some super-scalar capability. Research in this dimension led to development of most current *explicitly parallel instruction computer* (EPIC) architecture [2]. From what is made public to date it appears to be a wide instruction set architecture, with three instructions per 128-bit instruction parcel. It supports predicated and speculative execution. More over it can be seen that emphasis is made on employing dynamic scheduling technique in addition to the static scheduling.

## IV. INSTRUCTION PROCESSING

In a single threaded computer processing, the instruction takes place serially with nearly no hazards and stalls. However,

expanding the horizon towards pipelining makes a lot of achievement in speed, but also invites us to overcome a lot of interdependencies. Here the task of instruction processing is divided into several stages and processing of different instructions can be overlapped. By proper organization of the pipelining stages multiple instructions can be processed in one clock cycle and the updates that are valid is to be done in the last stage [1]. Determining how one instruction depends on the instructions can be made to flow through the pipeline like a serial program any updates that are needed to another is critical not only to the scheduling process but also to determine how much parallelism exits in a program and how that parallelism can be exploited. In particular to exploit instructional level parallelism we need to know which instructions can be executed in parallel. If two instructions are parallel they can execute simultaneously in a pipeline without causing stalls, assuming the pipeline has sufficient resources and hence no structural hazards exist. The main drawbacks that come in the development of ILP are pipeline hazards. Hazards are situations that prevent the next instruction in the instruction stream to execute during its designated clock cycle.

### A. Hazards and Dependencies

[1] provides a good classification of hazards into *Structural hazards,* one that arises due to limited resources, when hardware cannot support all possible combinations in simultaneous overlapped execution, *Data hazards* arise when an instruction depends on the previous instruction in a way that is exposed by overlapping of instructions in the pipeline and *Control hazards* arise from pipelining of branches and other instructions that change the program counter. Data Hazards can be considered as a separate class called *Dependencies* and this can be further classified into three groups Data dependency, Name dependency and Control dependency.

### B. IBM 360/91 machine overview

IBM 360/91, a pioneering machine of 1960s was responsible for several key contributions to high performance instruction processing [3]. This machine had four floating point registers and no cache. It took 12 cycles to service a load operation from memory. To tolerate this long latency, the load operation had to be overlapped with other instructions. With small number of available registers this meant possible overlap with other instructions that wrote into the same register as the load instruction. An important contribution of 360/91 was a mechanism of overcoming WAW hazards in instruction processing. This mechanism commonly known as Tomasulo's algorithm converts the sequential program representation into a data flow representation and then processes the instruction as per the ordering constraints in the dataflow representations rather than in the specified program order. WAW hazards as discussed earlier are overcome with register renaming. Here there are more physical storage elements that hold the results of the instruction than there are architectural registers. The architectural register names in an instruction can be used to obtain a name (i.e. a tag) for a physical storage element, and physical storage element names rather than architectural register names are used for instruction execution. This process becomes very important when the instructions are executed speculatively, regardless of the number of architectural registers. There physical registers will be needed to hold the outcomes of instructions to the correct instruction that creates the source operand value. The IBM 360/91 also recognized the importance of maintaining an uninterrupted supply of instructions for overall performance. To avoid a disruption when a branch instruction was encountered , it fetched instructions from both paths of a branch.[3] describes the entire organization of the machine IBM 360/91 with a final topic that deals with any design refinements which have been added to the basic organization.

### V. EXCEPTIONS

Interrupts are present in both in sequential architecture and in pipelined architecture. They cause the program to stop. However , when we come across the interrupt it is important for the architecture to note the process state, which generally contains the state of program counter, register and memory, so that the program can be started from the same point where it had stopped. In a pipelined architecture, depending on whether the state is saved or not we classify interrupts as *precise* or *imprecise* interrupts [4]. A precise interrupt is one that is consistent with the sequential architecture and reflects the following conditions, viz.,

*1) Instructions before program counter:* All instructions before the instruction indicated by the program counter are executed and have modified the process state correctly.

*2) Instructions after program counter:* All instructions after the instruction indicated by the program counter are not executed and have not modified the process state correctly.

*3) Interrupted Instruction handling:* If the interrupt is caused by the instruction exception, then the saved program counter should point to the interrupted instruction. The interrupted instruction may or may not be executed, depending on the definition of the architecture and cause of the interrupt.

If the saved process state is inconsistent with the sequential architectural model and does not satisfy the above conditions, then the interrupt is *imprecise*. Interrupts can also be classified as *Program interrupts*, that are caused due to the instructions and *External interrupts*, that are caused by sources outside the currently executing process, sometimes completely unrelated.

In 360/91 the interrupts were imprecise, because maintaining a precise state would require additional hardware and there were no reasons that would improve the performance of the machine significantly. It did not support virtual memory, where transparent handling of page fault exceptions is required, and therefore did not have one of the most important reasons for precise exception. The lack of precise exceptions had an important drawback that the machine was hard to debug, because the errors were not always reproducible. Precise interrupts were important when machines started arriving with virtual memory. Virtual memory is a transparent appearance of an addressable memory larger than what physically present in the machine [5]. This appearance of Virtual Memory, given by a layer of software, is responsible for moving pages between a virtual store and a physical store when a desired page is not present in the physical store. These software assume a sequential program execution, so to support transparent virtual memory, a precise machine state needs to be recovered at any point in the program at which page fault could occur.

### A. Implementing Precise Interrupts in CDC

[4] Describes schemes to implement precise interrupts, one of these schemes were used in the CDC Cyber 180.

The three methods are (1) Presence of a buffer in addition to the main register file, where in the buffer contains old values of registers and hence called "*History Buffer*", while the main register file is updated out of the program order. (2) *Reorder buffer*, this will contain all the instructions after they complete execution, it is responsible to reorder them as per the program order and then write it in register file. (3) a future file method, where there are two register files, an architecturally precise file that is updated in program order and an imprecise, or future, file that is updated out of the program order.

## VI. SPECULATIVE EXECUTION

Speculative Execution defines a methodology for avoiding constrains in parallelism by making some speculative early predictions. The main constraint that comes in parallelism is due to branching, because we do not know which instructions are to be executed till we know the out come of the branches. The main idea in Speculative Execution is that outcome of a branch is predicted and the instructions from the predicted path are executed. James E. Smith's paper on branch prediction techniques was the one that first described dynamic prediction schemes. The two bit counter described in the paper [6] was one of the most accurate counter. However the paper by Yeh and Patt introduced the concept of tracking the outcomes of previous branches and using this, along with the a history predictor as described by [6] . Further there was also a concept of control speculation by eliminating branches. This idea of conditional execution converted the control dependencies into data dependencies. Both of these techniques are being used today in the most recent EPIC architecture. There were some mechanisms like the Register Update Unit, suggested by the [8], where in a change in the model architecture is being used to facilitate precise interrupts and thus gives an assurance of smooth program execution. The paper [8] in its section 4.2 clearly lists the functions of the RUU. Another, mechanism in parallelism was demonstrated by [9] which basically dealt with developing an architecture that has all the requirements busy. [9] Introduced HPS (High Performance Substrate) for high performance computer engine, it also exposed a drawback or potential limitation of micro engine "One Operation per cycle". The paper emphasized on local parallelism as an alternative and selected restricted form of data flow based on the idea that parallelism available from the middle control flow tier (i.e. sequential control flow architecture) is highly localized. Their concept was to reduce the active instruction window, to exploit almost all the inherent parallelism in the program, thus reducing the synchronization cost that would be needed to keep the entire program around as a total data flow graph.

## VII. ARCHITECTURE OVERVIEW

### A. IBM RISC System/6000 Processor

[12] IBM RISC system/6000 processor is super-scalar processor that integrates packaging of chips in small packages. The general organization consists of an ICU that fetched instructions and executed the branch and condition register instructions. It dispatched two instructions per cycle to a Floating Point Unit (FPU), Fixed Point Unit (FXU). The FXU performed fixed-point arithmetic, address calculation for FP loads and stores, translates the addresses and controls the data cache. Thus the architecture allowed one floating-point, one fixed point, one branch and a LCR instruction to be performed simultaneously. One chip was dedicated in fetching multiple instructions in one cycle. The 2-way set associative instruction cache was composed of four small independent arrays that allowed four interleaving instructions to be fetched and placed in them computed by modulo 4. To avoid branching delays, four instructions are scanned ahead of time using the branch scanning logic so that the branch target can be fetched and outcome predicted. Branch prediction mechanisms are not used. When the target is evaluated, the pre-fetched instructions are either executed or discarded. Synchronization between the FPU uses instruction pre-fetch buffers IPB0-IBP3 that feeds data into decoders D0-D3. The FXU has a similar setup and uses PD0-PD3 as decoders. The FXU operations are always in R0 and R1 while FP uses other registers . Register renaming is used to prevent clashes between the FXU and FPU. R0 and R1 are re-name registers, and a map table entry, which is 32-entry, 6, bit wide maintains correspondence of an architectural register and the physical register. The Free List (FL) contains the list of unassigned registers, Pending –Target Return Queue (PTRQ) contains the physical registers used by instructions. Floating point uses BUSY and BYPASS registers to locate physical registers. This allows making use of each functional unit to the maximum provides synchronization between fixed-floating point units and precise interrupts can be implemented. This is a powerful, robust, high performance platform used for a wide range of applications.

### B. MIPS R10000 Superscalar Microprocessor

[13] MIPS R10000 is a dynamic super-scalar microprocessor implementing 64-bit Mips 4 instruction set architecture. It has one cluster bus connecting as many as 4 chips. The R10000 implements register mapping and non-blocking caches, which complement each other to overlap cache refill operations. Thus, if an instruction misses in the cache it must wait for its operand to be refilled, but other instructions can continue out of order. This increases the memory use and reduces effective latency, because refills begin early and up to four refills proceed in parallel while the processor executes other instructions. This type of cache design is called "non

blocking", because cache refills do not block subsequent accesses to other cache lines. The R10000 design includes complex hardware that dynamically reorders instruction execution based on operand availability. This hardware immediately adapts whenever cache misses delay instructions. The processor looks ahead up to 32 instructions to find possible parallelism. This instruction window is large enough to hide most of the refills from the secondary cache. However, it can hide only a fraction of main memory latency, which is typically much longer.

## VIII. FUTURE TRENDS AND DEVELOPMENT

As time flies, technology develops. Super-scalar architectures, where in, any program can be made parallel was an invention, then EPIC developed with its capabilities of using both predicted and control speculation for branches. The most recent invention in this filed is "Processor with conditional execution of every instruction", [10] is a patent filed on this invention on 16[th] April 2002. Here the author urges to consider all instructions as conditional instructions. He devices a scheme in which there is an instruction filed in each and every instruction. This instruction filed consists of a conditional identifier, which corresponds to a condition register. The condition register contains the condition value. If the condition value is the first condition, then the instruction is executed else the given instruction is treated as NOP if the said condition has the second condition value. Further if the condition identifier value is same as the value of the pre-selected identifier, then the instruction is unconditionally executed and not treated as a NOP. The author in this invention provides a general-purpose microprocessor architecture enabling more efficient computations of a type in which Boolean operations and arithmetic operations conditioned on the results of the Boolean operations are interleaved. The microprocessor is provided with a plurality of general purpose registers ("GPRs") and an arithmetic logic unit ("ALU"), capable of performing arithmetic operations and Boolean operations. The ALU has a first input and a second input, and an output, the first and second inputs receiving values stored in the GPRS. The output stores the results of the arithmetic logic unit operations in the GPRs. At least one of the GPRS is capable of receiving directly from the ALU a result of a Boolean operation. In one embodiment, at least one of the GPRS capable of receiving directly from the ALU a result of a Boolean operation is configured so as to cause the conditioning of an arithmetic operation of the ALU based on the value stored in the GPR. A method is also provided , performed in a microprocessor having such an architecture, in which a Boolean operation is performed in the ALU to obtain thereby a Boolean value representing the result of the Boolean operation. The Boolean value is stored in a first general purpose register in the same clock cycle as that in which the Boolean operation is performed. Thereafter, an arithmetic operation is performed in the arithmetic logic unit and the result of the arithmetic operation is stored in a second general purpose register. However, the step of performing/Storing is conditioned on the Boolean value stored in the first general purpose register. This concept on its self is an innovative one and further development is expected in this area. James Smith in his paper [11] describes a new trend in ILP. He says that sooner ILP will change into ILDP standing for instructional-level distributed processing, emphasizing inter-instruction communication with dynamic optimization and a tight interaction between hardware and low-level software. Now there will more considerations of the on-chip wire delays and power considerations will not permit billions of transistors to be used in a chip. ILDP micro-architecture paradigm described in this paper has distributed functional units, each fairly simple and with a very high frequency clock. Figure 1 below shows the ILDP micro-architecture. The communication between instructions should be localized to small units as this will take smaller wire as compared to having communication to longer units. The longer the wire, more is the delay. This type of unit is extremely power efficient and has minimal delay. This is where the future going to be in the recent years.

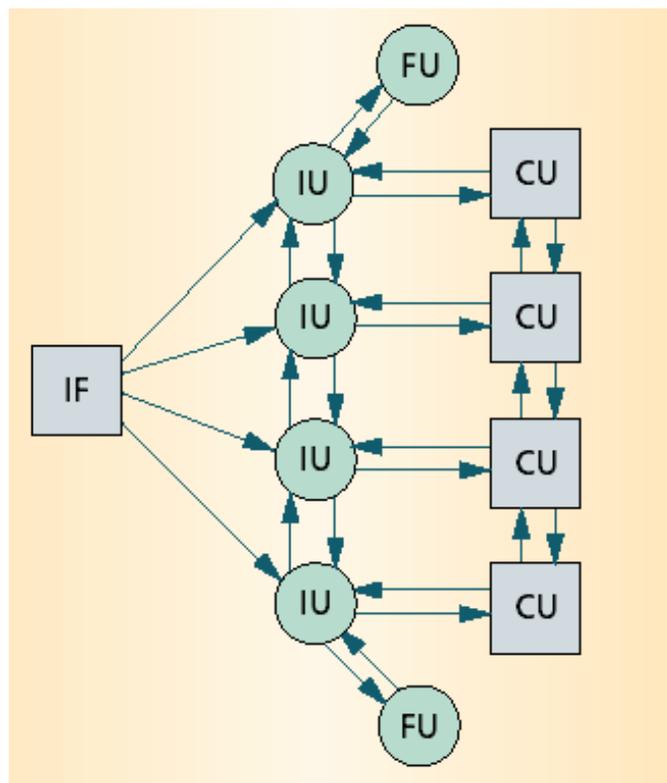

**ILDP Micro-architecture[11]**

## IX. CONCLUSION

This paper is a review of the papers in the Instruction Level Parallelism area. This survey of papers from the book, current literature and through recent patents in this area makes it clear that soon Instructional Level parallelism will be distributed, i.e. **Instruction Level Distributed Processing** (ILDP) and the exceptions will be removed by conditional prediction of each and every instruction. This will not only make the new computers power efficient but also minimize the delay to a large extent.